\documentclass[12pt,a4paper]{article}

\usepackage{graphicx}
\usepackage{cite}
\usepackage{bm}

\begin{document}
\begin{center}
\Large{\bf Role of doped Ru in coercivity-enhanced La$_{0.6}$Sr$_{0.4}$MnO$_{3}$ thin film studied by x-ray magnetic circular dichroism \\}

\vspace{12pt}

\small{T. Harano$^1$, G. Shibata$^1$, K. Ishigami$^2$, Y. Takashashi$^1$, V. K. Verma$^1$, V. R. Singh$^1$, \\
T. Kadono$^1$, A. Fujimori$^{1,\ 3}$, Y. Takeda$^3$, T. Okane$^3$, Y. Saitoh$^3$, H. Yamagami$^4$, T. Koide$^5$, \\
H. Yamada$^6$, A. Sawa$^6$, M. Kawasaki$^7$, Y. Tokura$^7$, A. Tanaka$^8$}

\vspace{12pt}

\normalsize{\it $^1$Department of Physics, The University of Tokyo, Bunkyo-ku, Tokyo 113-0033, Japan \\
$^2$Department of Complexity Science and Engineering, The University of Tokyo, Kashiwa-shi, Chiba 277-8561, Japan \\
$^3$Quantum Beam Science Directorate, Japan Atomic Energy Agency (JAEA), Soyo-gun, Hyogo 679-5148, Japan \\
$^4$Department of Physics, Kyoto Sangyo University, Kyoto-shi, Kyoto 603-8555, Japan \\
$^5$High Energy Accelerator Research Organization (KEK) Photon Factory, Tsukuba, Ibaraki 305-0016, Japan \\
$^6$National Institute of Advanced Industrial Science and Technology (AIST), Tsukuba, Ibaraki 305-8562, Japan \\
$^7$Department of Applied Physics, The University of Tokyo, Bunkyo-ku, Tokyo 113-0033, Japan \\
$^8$Department of Quantum Matter, ADSM, Hiroshima University, Higashi-Hiroshima-shi, Hiroshima 739-8530, Japan \\}

\end{center}

\section*{Abstract}
The coercivity of La$_{1-x}$Sr$_x$MnO$_3$ thin films can be enhanced by Ru substitution for Mn. In order to elucidate its mechanism, we performed soft x-ray absorption and magnetic circular dichroism measurements at the Ru M$_{2,3}$ and Mn L$_{2,3}$ edges. We found that the spin direction of Ru and Mn are opposite and that Ru has a finite orbital magnetic moment. Cluster-model analysis indicated that the finite orbital magnetic moment as well as the reduced spin moment of Ru result from local lattice distortion caused by epitaxial strain from the SrTiO$_3$ substrate in the presence of spin-orbit interaction.

\vspace{24pt}

Hole-doped manganese oxides with the perovskite-type structure have been known as materials in which interplay between the spin, orbital, and charge degrees of freedom leads to various phases and functionalities \cite{JonkerPhysica1950, TokuraGMR, TokuraScience2000}. Among them, La$_{1-x}$Sr$_{x}$MnO$_{3}$ (LSMO) with 0.1 $\leq$ {\it x} $\leq$ 0.5 is a half-metal which showing giant magneto-resistance \cite{PD_LSMObulk}. LSMO thin films have high potential for spintronics applications but there is a serious problem that the coercivity ({\it H}$_{\rm C}$) is too small: less than 10 Oe at 300 K. Yamada {\it et al.} \cite{Yamada} reported that one can enhance the coercivity of LSMO thin films by substituting Ru for Mn. They also observed a decrease of the magnetization with Ru doping, and attributed this to antiferromagnetic coupling between Mn and Ru. They proposed that charge transfer occurs from Mn$^{4+}$ to Ru$^{4+}$ (Mn$^{4+}$+Ru$^{4+}$ $\to$ Mn$^{3+}$+Ru$^{5+}$). In order to elucidate the mechanism of the coersivity enhancement, one needs microscopic information about the Ru atom such as the valence, spin, and orbital  states in the LSMO matrix.


X-ray absorption spectroscopy (XAS) and x-ray magnetic circular dichroism (XMCD) are powerful element-specific probes of the electronic and magnetic properties of complex materials. XMCD is the difference between XAS taken with light of positive and negative helicities. In this Letter, we have performed XAS and XMCD measurements at the Ru M$_{2,3}$ and Mn L$_{2,3}$ absorption edges of Ru-doped La$_{0.6}$Sr$_{0.4}$MnO$_{3}$ thin film in order to investigate the electronic and magnetic states of each element, in particular of Ru, and to elucidate the mechanism of the coercivity enhancement. 


We fabricated a Ru-doped La$_{0.6}$Sr$_{0.4}$MnO$_{3}$ thin film by the plused laser deposition (PLD) method on a (001)-oriented SrTiO$_{3}$ (STO) substrate. 5 $\%$ of Mn atoms were replaced by Ru atoms, and therefore, the composition was La$_{0.6}$Sr$_{0.4}$Mn$_{0.95}$Ru$_{0.05}$O$_{3}$. The thickness of this sample was 50 nm. The details of the sample fabrication were described elsewhere \cite{Yamada}. XAS and XMCD measurements were performed at BL23-SU of SPring-8. The spectra were recorded in the total electron yield (TEY) mode at the temperature {\it T} = 10 K, well below the Curie temperature {\it T}$_{\rm C}$ $\sim$ 355 K,  and in the magnetic field {\it H} = 3 T applied perpendicular to the film surface. In order to detect the weak XMCD signals of the Ru M$_{2, 3}$ edges, we used the helicity-switching mode operated at 1 Hz \cite{1Hz}.

\begin{figure}[b]
\begin{center}
\includegraphics[width=12cm]{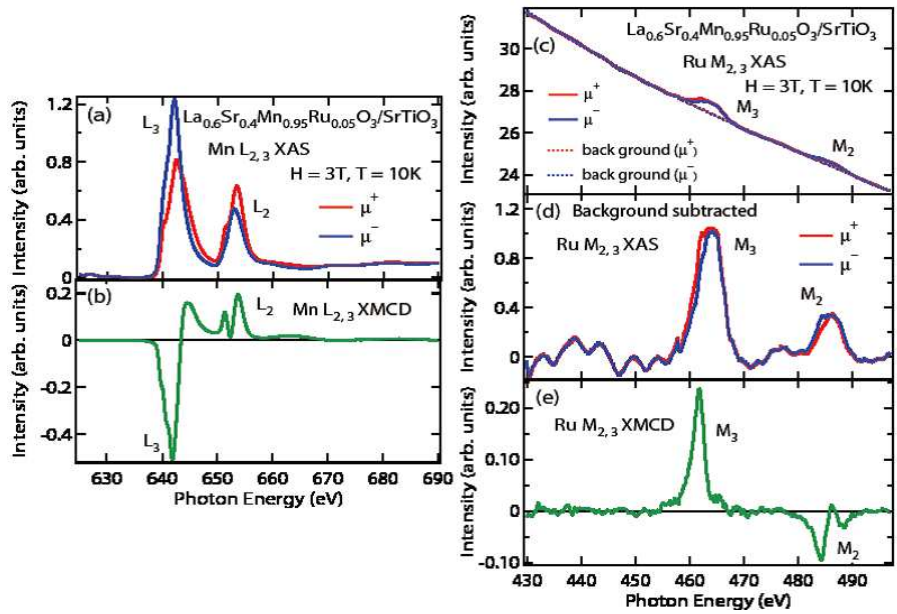}
\caption{(Color online) (a), (b) XAS and XMCD spectra at the Mn L$_{2,3}$ edge of La$_{0.6}$Sr$_{0.4}$Mn$_{0.95}$Ru$_{0.05}$O$_{3}$ thin film under the magnetic field of {\it H} = 3 T at temperature {\it T} = 10 K.
(c), (d), (e) Corresponding XAS and XMCD spectra at the Ru M$_{2,3}$ edge. (c) shows raw data and (d) shows data with the smooth background subtracted.}
\label{Fig123}
\end{center}
\end{figure}

The XAS and XMCD spectra at the Mn L$_{2, 3}$ edge are shown in Figs. \ref{Fig123} (a) and (b), respectively. The spectra well agree with those of a pure La$_{0.6}$Sr$_{0.4}$MnO$_{3}$ bulk sample reported in Ref. \cite{KoideLSMObulk}. The XAS and XMCD spectra at the Ru M$_{2, 3}$ edge are shown in Figs. \ref{Fig123}  (c), (d), and (e). They are the raw Ru M$_{2, 3}$ XAS data, the XAS spectra with the back ground subtracted, and the XMCD spectrum, respectively.
The XMCD intensity was recorded at each photon energy in the helicity switching mode, which enabled us to detect the weak XMCD signals at the Ru M$_{2,3}$ edge. The XMCD spectrum was then averaged between the two magnetic directions parallel and antiparallel to the photon propagation vector.
From the opposite signs of the XMCD signals between the Mn L$_{2, 3}$ and Ru M$_{2, 3}$ edges, one can unambiguously conclude that the spin direction of Ru is opposite to that of Mn, consistent with Yamada {\it et al.}'s observation of the decreased magnetization upon Ru doping \cite{Yamada}.
We have applied the XMCD sum rules \cite{TholePRL1992, CarraPRL1993} to the measured spectra and deduced the spin and orbital magnetic moment of Mn and Ru as summarized in Table \ref{spin_orbital_m_Ru-LSMO_and_SRO}. One can see that the orbital moments of Ru is finite and parallel to the spin moment in contrast to the negligibly small orbital magnetic moment in bulk SrRuO$_{3}$ (SRO) \cite{OkamotoPRB}. 

\begin{figure}[t]
\begin{center}
\includegraphics[width=8cm]{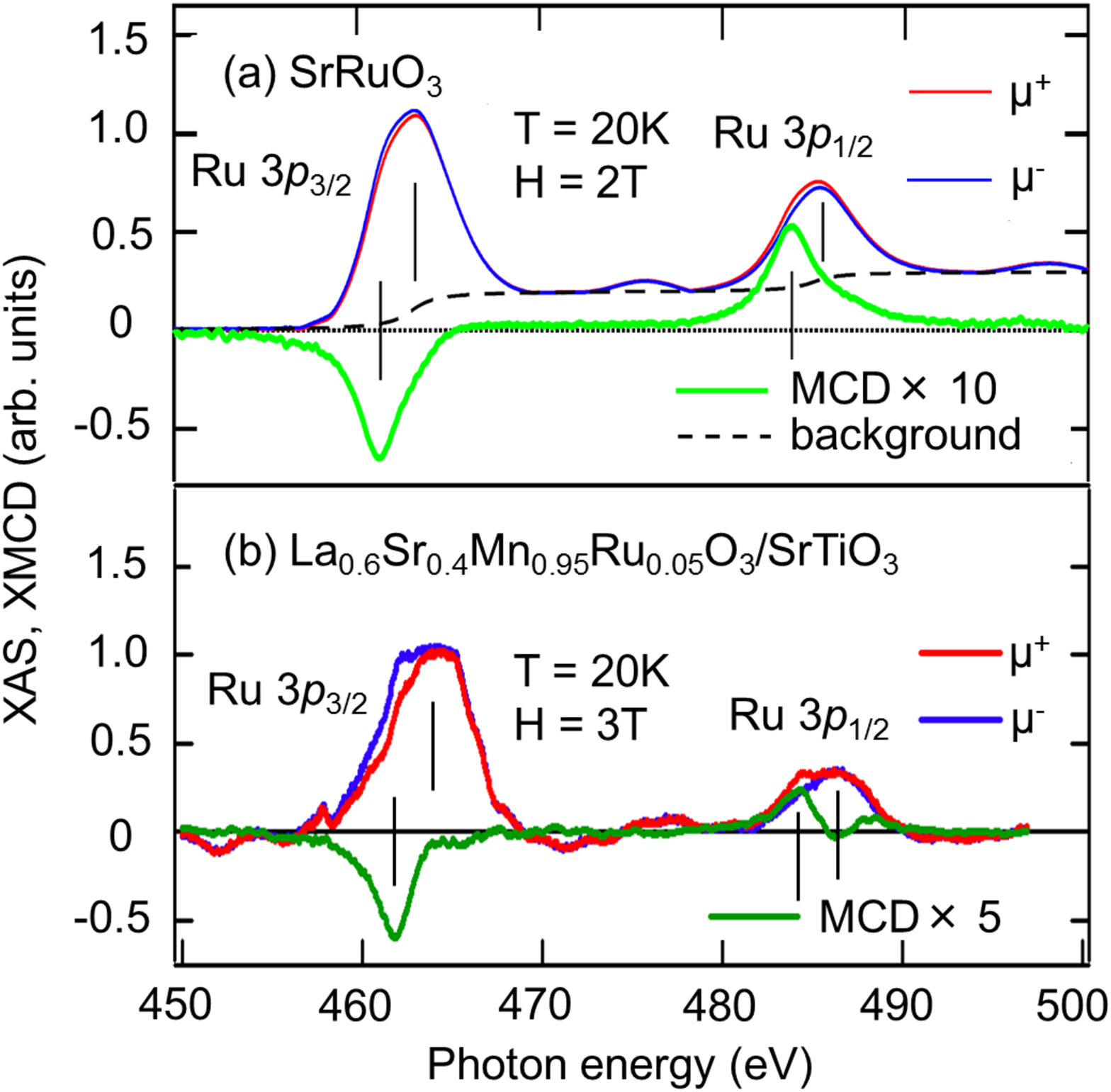}
\caption{(Color online) XAS and XMCD spectra at the Ru M$_{2,3}$ edge of bulk SrRuO$_{3}$ polycrystal \cite{OkamotoPRB} (a) compared with those of the Ru-doped LSMO thin film (b).}
\label{SRORuLSMO}
\end{center}
\end{figure}

Figures \ref{SRORuLSMO} (a) and (b) show comparison of the XAS and XMCD spectra of Ru-doped LSMO with those of bulk SRO. In the case of Ru-doped LSMO, one can recognize structures which can be attributed to the unoccupied {\it e}$_{g}$ and {\it t}$_{2g}$ orbitals of the Ru 4{\it d} manifold. Corresponding features are seen in the spectra of SRO, which means that the {\it e}$_{g}$-{\it t}$_{2g}$ crystal-field splitting of Ru-doped LSMO is similar to that of SRO. The spin and orbital magnetic moments of Ru-doped LSMO and SRO are indicated in Table \ref{spin_orbital_m_Ru-LSMO_and_SRO} \cite{OkamotoPRB}. For Ru-doped LSMO, the spin magnetic moment {\it m}$_{\rm spin}$ = 0.64 {\it $\mu$}$_{\rm B}$ is as small as that ({\it m}$_{\rm spin}$ = 0.6 {\it $\mu$}$_{\rm B}$) of SRO but the orbital magnetic moment {\it m}$_{\rm orb}$ = 0.14 {\it $\mu$}$_{\rm B}$ is much larger than that ({\it m}$_{\rm orb}$ = 0.04 {\it $\mu$}$_{\rm B}$) of SRO. As we shall see below, the values of {\it m}$_{\rm spin}$ and {\it m}$_{\rm orb}$ sensitively depend on the strengths of spin-orbit interaction and epitaxial strain. 

\begin{figure}[t]
\begin{center}
\includegraphics[width=12cm]{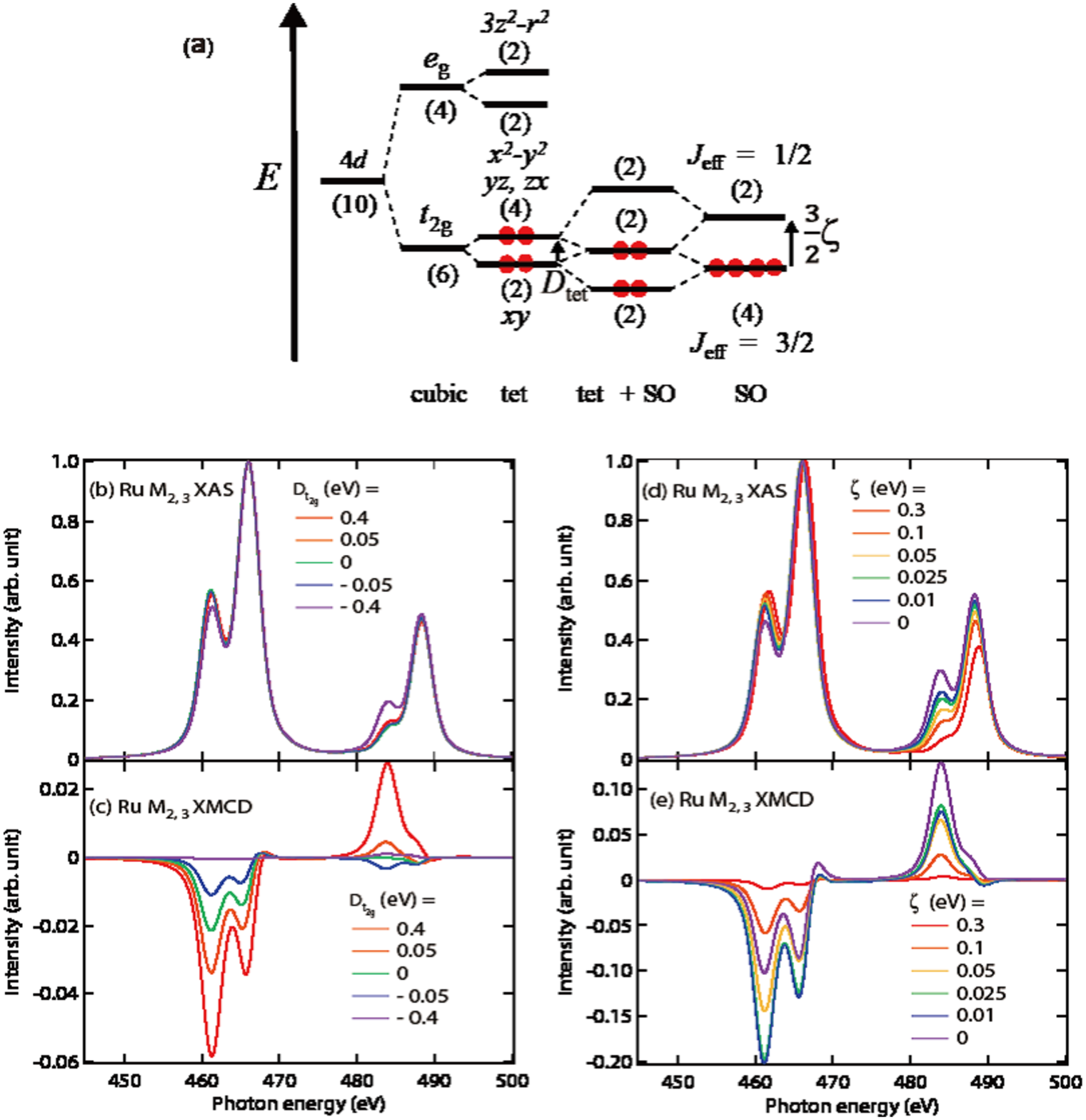}
\caption{(Color online) Cluster-model calculations of the XAS and XMCD spectra at the Ru M$_{2,3}$ edge for the Ru$^{4+}$ valence state. (a) One-electron energy-level diagram of the Ru$^{4+}$ ion in the presence of  both spin-orbit (SO) interaction and tetragonal lattice distortion, which split the $t_{2g}$ level by $D_{t_{2g}}$. The degeneracy of each level is indicated in a bracket. (b), (c) Tertragonal crystal-field ({\it D}$_{t_{2g}}$) dependence of the spectra. Spin-orbit parameter {\it $\zeta$} is fixed at 0.1eV and the other parameters are given in Table \ref{parameta}. (d), (e) Spin-orbit interaction ($\zeta$) dependence of the spectra. {\it D}$_{t_{2g}}$ is fixed at 0.4 eV.}
\label{egSO}
\end{center}
\end{figure}

Tetragonal local lattice distortion around the Ru atom is expected to be present for the epitaxial thin film \cite{KonishiJPSJ} and to split the {\it t}$_{2g}$ level of Ru into the {\it d}$_{xy}$ and {\it d}$_{zx}$/{\it d}$_{yz}$ sub-levels as shown in Fig. \ref{egSO} (a). Here, the splitting of the {\it t}$_{2g}$ level is denoted by {\it D}$_{t_{2g}}$. (For the splitting of the $e_g$ level, $D_{e_g}/D_{t_{2g}} =$ 2 has been assumed.) The spin-orbit coupling constant of of the Ru 4d orbitals is {\it $\zeta$} = 0.1 eV \cite{MizokawaPRL,Herman}, which is  larger than those ($\sim$0.01 eV) of the 3{\it d} orbitals of the first-series transition-metal atoms. {\it $\zeta$} also causes the energy splitting between the {\it J}$_{\rm eff}$ = 1/2 and 3/2 sub-levels \cite{Kim06032009}. If there were no tetragonal crystal field, the spin magnetic moment would be more dramatically reduced because four 4{\it d} electrons occupy the {\it J}$_{\rm eff}$ = 3/2. In the presence of the tetragonal crystal field, the ground state becomes a mixture of {\it J}$_{\rm eff}$ = 1/2 and 3/2 states and has finite spin and orbital magnetic moments. In order to interpret the {\it m}$_{\rm spin}$ and {\it m}$_{\rm orb}$ of Ru doped in LSMO, we have performed cluster-model calculations on the Ru M$_{2, 3}$-edge XAS and XMCD spectra. Other parameters and their values used for the calculation are taken from Ref. \cite{parameta} and listed in Table \ref{parameta}. (Because multiplet splitting became too large if the tabulated Slater integrals for the Ru 4{\it d} and 3{\it p} orbitals \cite{Mann} were used, 25 \% of Sleater integrals were used representing the strong {\it p-d} hybridization and/or the delocalization of 4{\it d} electrons in 4{\it d} transition-metal oxides.) The XAS and XMCD spectra for Ru$^{4+}$ calculated for various parameter sets are shown in Fig. \ref{egSO} (b) to (e). In panels (b) and (c), the {\it D}$_{t_{2g}}$ dependence of the spectra with fixed $\zeta$ = 0.1 eV is shown. Spectra for {\it D}$_{t_{2g}}$ = 0.4 eV  best reproduce the XAS and XMCD spectra concerning the line shapes and the {\it m}$_{\rm spin}$ and {\it m}$_{\rm orb}$ values.
(The splitting of the M$_{3}$ edge, however, could not be eliminated within the physically reasonable parameter range, probably because the {\it e}$_{g}$ orbitals of Ru is too itinerant to properly describe using the cluster model.)
If we change {\it $\zeta$}, agreement with experiment becomes worse as shown in Figs. \ref{egSO} (d) and (e).

\begin{figure}[t]
\begin{center}
\includegraphics[width=7.8cm]{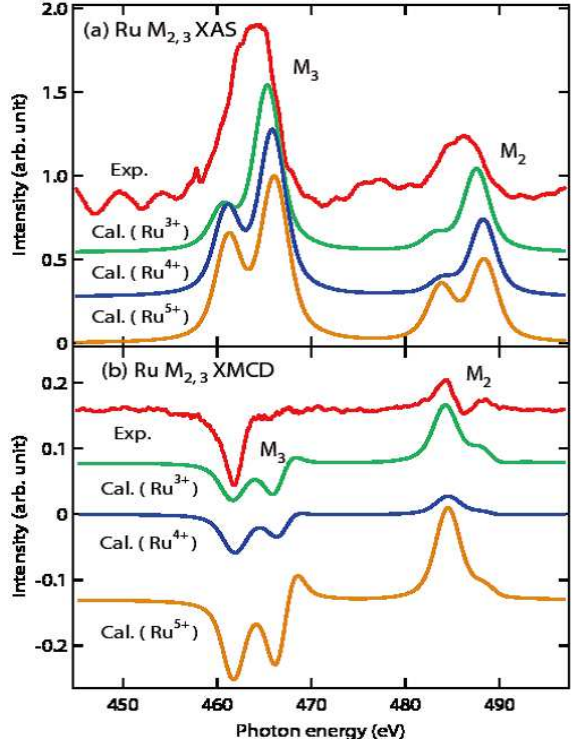}
\caption{(Color online) 
Comparison of the experimental Ru M$_{2,3}$ XAS (a) and XMCD (b) spectra of La$_{0.6}$Sr$_{0.4}$Mn$_{0.95}$Ru$_{0.05}$O$_{3}$ with cluster-model calculation for the various valence states of Ru$^{3+}$, Ru$^{4+}$, and Ru$^{5+}$. The calculation for Ru$^{4+}$ is in better agreement with experiment than Ru$^{5+}$ and Ru$^{3+}$. If Ru$^{5+}$ is assumed, {\it m}$_{\rm spin}$ becomes too large and therefore the XMCD spectra becomes much stronger than experiment. If Ru$^{3+}$ is assumed, the direction of the orbital magnetic moment becomes opposite to the spin magnetic moment, in disagreement with experiment.}
\label{Fig56}
\end{center}
\end{figure}

The observed small spin ($m_{\rm spin}=$ 0.64 {\it $\mu$}$_{\rm B}$) and finite orbital magnetic moment ($m_{\rm orb}=$0.14 {\it $\mu$}$_{\rm B}$) of the Ru$^{4+}$ ion can thus be attributed to  the combined effect of the epitaxial strain and spin-orbit interaction of the Ru$^{4+}$ ({\it d}$^{  4}$) ion. In order to test the possibilities of Ru valences other than 4+, we performed calculations for Ru$^{3+}$ and Ru$^{5+}$, too, as shown in Figs. \ref{Fig56} (a) and (b), but no better agreement with experiment could be obtained: If Ru$^{5+}$ is assumed, {\it m}$_{\rm spin}$ becomes too large (Table \ref{spin_orbital_m_Ru-LSMO_and_SRO}) and therefore the XMCD spectra becomes much stronger than experiment. If Ru$^{3+}$ is assumed, the direction of the orbital magnetic moment becomes opposite to the spin magnetic moment, in disagreement with experiment (Table \ref{spin_orbital_m_Ru-LSMO_and_SRO}). 


The positive {\it D}$_{t_{2g}}$ value deduced in the present study indicates that the RuO$_{6}$ octahedron is compressed within the {\it a-b} plane. At first glance, this appears contradictory to the tensile strain of the LSMO thin film grown on STO substrate \cite{KonishiJPSJ}. However, because the radius of Ru$^{4+}$ ion is larger than the average radius of the Mn$^{3+}$/Mn$^{4+}$ ion as well as the radius of the Ti$^{4+}$ ion in the STO substrate \cite{Shannon}, the RuO$_{6}$ octahedron in the LSMO thin film epitaxially grown on the STO substrate will be compressed within the {\it a-b} plane while it may be elongated along  the flexible {\it c}-direction.
In fact, SrRuO$_3$ thin films grown on SrTiO$_3$ substrates exhibit easy magnetization axis perpendicular to the film \cite{xia}, like the appearance of orbital magnetic moment in the present case.
We therefore consider that interplay between the epitaxial strain and spin-orbit interaction gives rise to the finite orbital magnetic moment of Ru$^{4+}$, which enhances the magnetic anisotropy. 
Since magnetic anisotropy generally ehnaces the hysteretic behavior of magnetization and hence the coercivity, the coersivity of the LSMO thin film is enhanced by Ru doping. 


In conclusion, we have performed XAS and XMCD studies of the Ru-doped LSMO thin film in order to investigate the origin of the enhanced coercivity. From the XAS and XMCD spectra, it has been found that the spin direction of Ru is opposite to that of Mn and that Ru has a finite orbital magnetic moment unlike that in SRO. On the basis of the cluster-model analysis of the spectra, we conclude that the valence of Ru is 4+ and the finite orbital magnetic moment arises from the spin-orbit interaction and tetragonal crystal field arising from the epitaxial strain from the STO substrate. Such a Ru atom would be magnetically anisotropic and hence enhance the coercivity of the Ru-doped LSMO thin films. In order to confirm this scenario, measurement of the orbital magnetic moment for different magnetic field directions as well as measurements of hysteresis using the element-specific XMCD technique are desired.


We would like to thank J.-M Chen for informing us of the cluster-model calculations on the Ru 2{\it p} core-level spectra of LaCe$_{0.5}$Ru$_{0.5}$O$_{3}$ prior to publication.  
This work was supported by a Grant-in-Aid for Scientific Research from JSPS (S22224005) and the Quantum Beam Technology Development Program from JST.
The experiment was performed under the approval of the Photon Factory Program Advisory Committee (proposal No. 2010G187) and under the Shared Use Program of JAEA Facilities (Proposal No. 2011A3840, No. 2012A3824/BL23SU).

\clearpage

\clearpage

\clearpage

\clearpage

\renewcommand{\thetable}{\Roman{table}}

\begin{table}[t]
\begin{center}
\caption{Spin and orbital magnetic moments ({\it m}$_{\rm spin}$, {\it m}$_{\rm orb}$) of Mn and Ru in La$_{0.6}$Sr$_{0.4}$Mn$_{0.95}$Ru$_{0.05}$O$_{3}$ (Ru-doped LSMO) in units of {\it $\mu$}$_{\rm B}$ deduced from the XAS and XMVCD spectra using the sum rules. The deduced values are  compared with those derived from the cluster-model calculation for the Ru$^{3+}$, Ru$^{4+}$, and Ru$^{5+}$ valence states. The {\it m}$_{\rm spin}$ and {\it m}$_{\rm orb}$ of  SrRuO$_{3}$ (SRO) \cite{OkamotoPRB} are also indicated for comparision. The spin directions of Mn and Ru are taken to be positive and negative, respectively, corresponding to those in Ru-doped LSMO. }
\vspace{1cm}
\begin{tabular}{ccccccc}
\hline
& \multicolumn{2}{c}{Expt.} & \multicolumn{3}{c}{Cluster-model} & Expt. \\
& \multicolumn{2}{c}{Ru-doped LSMO} & \multicolumn{3}{c}{Calculation} & SRO \cite{OkamotoPRB} \\
& Mn & Ru & Ru$^{3+}$ & Ru$^{4+}$ & Ru$^{5+}$ & Ru \\ \hline
{\it m}$_{\rm spin}$ & 3.58 & -0.64 & -0.86 & -0.50 & -2.7 & -0.6 \\
{\it m}$_{\rm orb}$ & 0.02 & -0.14 & 0.98 & -0.23 & 0.08 & -0.04 \\ \hline
\label{spin_orbital_m_Ru-LSMO_and_SRO}
\end{tabular}
\end{center}
\end{table} 

\clearpage

\begin{table} [t]
\begin{center}
\caption{Electronic structure parameters used for the cluster-model calculation of the Ru M$_{2,3}$ XAS and XMCD spectra for Ru$^{3+}$, Ru$^{4+}$, and Ru$^{5+}$ in units of eV. 10{\it Dq}: Octahedral crystal-field splitting, {\it $\Delta$}: {\it p}-to-{\it d} charge-transfer energy, {\it U}$_{4d-4d}$: On-site 4{\it d}-4{\it d} Coulomb energy, {\it U}$_{4d-3p}$: On-site 4{\it d}-3{\it p} core Coulomb energy, {\it pd}{\it $\sigma$}: Hopping integral between the 3{\it d} and oxygen {\it p} orbitals (Slater-Koster parameter). These parameter values are taken from Ref. \cite{parameta}.}
\vspace{1cm}
\begin{tabular}{cccccc}
\hline
& 10{\it Dq} & {\it $\Delta$} & {\it U}$_{4d-4d}$ & {\it U}$_{4d-3p}$ & {\it pd}{\it $\sigma$}  \\ \hline
Ru$^{3+}$ & 3.5 & 5 & 3 & 3.6 & -2.31 \\
Ru$^{4+}$ & 3.5 & 2 & 3 & 3.6 & -2.52 \\
Ru$^{5+}$ &3.5 & -1 & 3 & 3.6 & -2.78 \\ \hline
\label{parameta}
\end{tabular}
\end{center}
\end{table} 

\clearpage

\graphicspath{{Ru-LSMO/fig/}}

\end{document}